# Generalized Brewster-angle effect in thin-film optical absorbers and its application for graphene hydrogen sensing

Kandammathe Valiyaveedu Sreekanth [a, b$], Mohamed ElKabbash[c, d$], Rohit Medwal[e], Jihua Zhang[c], Theodore Letsou[d], Giuseppe Strangi[d, f], Michael Hinczewski[d], Rajdeep S. Rawat[e], Chunlei Guo[*c,] and Ranjan Singh[*a, b]

[a.]Division of Physics and Applied Physics, School of Physical and Mathematical Sciences, Nanyang Technological University, 21 Nanyang Link, Singapore-637371

[b.]Centre for Disruptive Photonic Technologies, The Photonic Institute, 50 Nanyang Avenue, Singapore-639798

[c.]The Institute of optics, University of Rochester, 275 Hutchison Rd, Rochester, NY 14620, USA

[d.]Department of Physics, Case Western Reserve University, 10600 Euclid Avenue, Cleveland, OH, 44106 (USA)

[e.]Natural Sciences and Science Education, National Institute of Education, Nanyang Technological University, Singapore 637616, Singapore

[f.]Centre for Advanced 2D Materials and Graphene Research Centre, National University of Singapore, Singapore, 117542.

[j.]CNR-NANOTEC and Department of Physics, University of Calabria, 87036 - Rende (Italy)

**AUTHOR INFORMATION**

[$]These authors (K. V. Sreekanth and M. ElKabbash) contributed equally to this work.

* E-mail: ranjans@ntu.edu.sg (Ranjan Singh)

* E-mail: guo@optics.rochester.edu (Chunlei Guo)

**ABSTRACT**

Generalized Brewster angle (GBA) is the incidence angle at polarization by reflection for *p*- or *s*-polarized light takes place. Realizing *s*-polarization Brewster effect requires a material with magnetic response which is challenging at optical frequencies since the magnetic response of materials at these frequencies is extremely weak. Here, we experimentally realize GBA effect in the visible using a thin-film absorber system consisting of a dielectric film on an absorbing substrate. Polarization by reflection is realized for both *p*- and *s*- polarized light at different angles of incidence and multiple wavelengths. We provide a theoretical framework for the generalized Brewster effect in thin-film light absorbers. We demonstrate hydrogen gas sensing using a single layer graphene film transferred on a thin-film absorber at the GBA with ~1 fg/mm$^2$ aerial mass sensitivity. The ultrahigh sensitivity stems from the strong phase sensitivity near point of darkness, particularly at the GBA, and the strong light-matter interaction in planar nanocavities. These findings depart from the traditional domain of thin-films as mere interference optical coatings and highlight its many potential applications including gas sensing and biosensing.

**KEYWORDS**: generalized Brewster angle, thin film optical absorbers, visible frequencies, graphene, hydrogen sensing

The Brewster angle, $\Theta_B$ is commonly defined as the angle at which the Fresnel's reflection coefficients for *p*-polarized light vanishes[1]. For a given medium, the sum of the incident angle $\Theta_i$ and transmitted angle $\Theta_t$ is $\pi/2$ when $\Theta_i = \Theta_B$. The orthogonality condition draws a simple picture for realizing the Brewster angle effect. Light incident on a medium at $\Theta_B$ induces electron oscillations in the direction of the electric field which give rise to reflected wavelets. The oscillating electrons, however, do not produce a field at points on the axis of oscillation. When $\Theta_i + \Theta_t = \pi/2$, the oscillation axis is in the direction of the reflected wave, hence, no reflection takes place. The Brewster effect is widely used in different applications, e.g., reducing glare of sun reflecting off horizontal surfaces and in laser physics in gain media, cavity mirrors, and in prisms to minimize reflection losses, as well as in high performance terahertz modulators[2]. Furthermore, Brewster angle microscopes rely on the enhanced phase sensitivity near the Brewster angle and can image monolayers at the air-liquid interface[3-4]. However, the phase sensitivity at the Brewster angle is thought to be insufficient for sensing applications as the local electric fields for dielectric substrates are small[4].

The aforementioned common definition of the Brewster angle, however, assumes that the reflection occurs from a plane wave incident on a homogenous, non-magnetic, achiral, and isotropic material. The Brewster effect can take place for either *p*- or *s*- polarized light if one of these assumptions is violated. For example, it is known that a Brewster angle for *s*-polarized light exists in a magnetic material with permeability $\mu \neq 1$. In magnetic materials, there exists an angle where the reflected *s*-polarized light vanishes $\Theta_B^s$, and another angle where the reflected *p*-polarized light vanishes $\Theta_B^p$, such that $\Theta_B^s \neq \Theta_B^p$ for non-normal incidence [5]. For a magnetic medium, the sum of the incident angle and the GBAs $\Theta_B^{p,s}$ does not need to be $\pi/2$ as long as total destructive interference between the magnetic and electric dipoles takes place at $\Theta_B^{p,s}$. Accordingly, the Brewster effect can take place even when the oscillation axis is not parallel to the wave reflection direction and the wavelets produced by individual oscillating electrons do not vanish at the reflection direction. What is necessary, however, is that the vector sum of the radiated field vanishes in the reflection due to destructive interference[6]. The Brewster angle, in its most general form, is the angle where *only* a single polarization is reflected due to the destructive interference between radiating electric and/or magnetic dipoles for the orthogonal polarization.

Realizing generalized Brewster effect at optical frequencies is challenging since the magnetic response of materials at these frequencies is very weak, i.e., $\mu \sim 1$. Metamaterials, however, can support negative permeability, thus a magnetic response is possible[7-9]. The generalized Brewster effect was realized experimentally using split ring resonators in the microwave region[10-11]. In the optical regime, GBA effect was demonstrated using all-dielectric metamaterials[12]. In addition to the intense lithography required to fabricate a metamaterial with a magnetic response, the Brewster angle demonstrated did not realize complete polarization of the reflected light in the visible frequencies, particularly for *s*-polarized light[12]. Strictly speaking, however, this demonstration did not exhibit true Brewster effect, rather showed unequal reflection for *s*- and *p*-polarizations which is a natural consequence of the Fresnel equations even in the absence of any magnetic response. On the other hand, *s*-polarized Brewster effect was shown in stratified metal-dielectric metamaterials due to changes in the effective magnetic permeability of the thin-film stack[13]. In addition, for nonmagnetic media, the *s*-polarized Brewster effect was demonstrated by adding a two-dimensional material at the interface between two media when the conditions for total internal reflection are satisfied such that reflected *s*-polarized light is absorbed fully by the graphene layer[14]. Furthermore, GBA was demonstrated in anisotropic materials[15] and chiral materials[16].

The assumption that the reflection occurs from a homogenous medium can also be violated by creating a multilayer structure where either the *s*- or the *p*- polarized light is reflected and the other polarization is extinguished[17]. The generalized Brewster conditions of lithography-free planar stack of thin-films have been theoretically investigated where the inhomogeneity is due to stacking different materials[18-24]. In this case, the GBA corresponds to an angle where electric dipoles in the inhomogeneous stack of materials destructively interfere. An experimental realization of GBA effect of a transparent film on an absorbing substrate, however, has not been demonstrated. Furthermore, the realization of the generalized Brewster effect can be used for sensing applications providing that it is associated with strong field localization.

In this letter, we investigate theoretically and experimentally the generalized (*p*- and *s*- polarization) Brewster conditions of a lossless dielectric film on an absorptive substrate at multiple wavelengths in the visible spectral region. By demonstrating, thin-film interference based perfect light absorption of a single polarization, the Brewster effect, i.e., polarization by reflection, is realized for both *s*- and *p*- polarized light at different angles of incidence. We further demonstrate hydrogen sensing using a hybrid platform of single layer graphene and the thin-film absorber. The realization of phase singularity in the ellipsometry phase parameter at the GBA accompanied by strong field confinement within the graphene layer in the thin-film cavity enabled ultrahigh hydrogen sensitivities of ~ 1 fg/mm$^2$ with cheap materials and scalable fabrication process.

**Theory of Generalized Brewster effect in thin-film light absorbers**

We investigate the proposed design, i.e., a lossless dielectric film on a substrate with optical losses. Our system consists of a superstrate (refractive index $n_0$), a dielectric layer (refractive Index $n_d$, thickness d), and a lossy substrate (refractive index $n_s + ik_s$). Using transfer matrix theory[25], we obtain expressions for the conditions necessary to realize the GBA effect for *p*- and *s*- polarized light in terms of the incident angle $\theta_0$ and the phase thickness of the dielectric layer $\Phi_d \equiv 2\pi d\lambda^{-1}\sqrt{n_d^2 - n_0^2 \sin^2(\theta_0)}$ (see Supporting Information for detailed derivation).

(i) *p*-polarization:

$$\theta_0 \approx \tan^{-1}\left[\frac{n_s}{n_0}\left(1 + \frac{1}{2}k_s^2\left(\frac{n_0^2(n_0^2 - 3n_d^2)}{(n_0^2 - n_d^2)n_s^4} + \frac{2(n_0^2(2n_d^2 - 3n_s^2) + n_s^4)}{n_s^2(n_s^2 - n_0^2)(n_s^2 - n_d^2)}\right)\right)\right], \quad (1)$$

$$\tan(\Phi_d) \approx \frac{n_d^2 k_s (n_s^2 - n_0^2)\sqrt{n_0^2(n_d^2 - n_s^2) + n_d^2 n_s^2}}{(n_0^2 - n_d^2)n_s^3(n_s^2 - n_d^2)}$$

The first condition above defines a unique GBA $\theta_0$ at which *p*-polarized reflection is zero. Note that when $k_s \to 0$ it reduces to the standard Brewster angle, $\theta_0 \to \tan^{-1}(n_s/n_0)$. For finite $k_s$ with the materials we use, the correction due to a lossy substrate is quite small, so $\theta_0$ remains close to the conventional Brewster angle. The $\tan(\Phi_d)$ condition can be solved to find a set of dielectric layer thicknesses *d* that will give zero reflection (there is more than one possibility since $\tan(\Phi_d + m\pi) = \tan(\Phi_d)$ for any integer *m*. Note that when $k_s \to 0$, this condition reduces to $\tan(\Phi_d) = 0$. In this case, one possible solution is *d* = 0, the conventional case where no dielectric layer is present. In addition, the conventional Brewster effect ($\theta_0 = \tan^{-1}(n_s/n_0)$) is realized for other *d* values that satisfy the $\tan(\Phi_d) = 0$. However, when $k_s > 0$ we need a finite *d* > 0 to achieve zero reflection, i.e., there is no Brewster angle for a dielectric with optical losses unless we add an additional lossless dielectric with finite thickness.

(ii) *s*-polarization:

To get zero reflection for *s*-polarization, the real parts of the refractive indices must satisfy $n_0 n_s > n_d$. Additionally, to get compact expressions, we assume $n_s > n_d > n_0$. The conditions are then given by:

$$\theta_0 \approx \tan^{-1}\left[\sqrt{\frac{n_0^2 n_s^2 - n_d^4}{(n_0^2 - n_d^2)^2}\left(1 + k_s^2 \frac{n_0^2(n_d^4 - (n_0^2 - 2n_d^2)n_s^2 - 2n_s^4)}{2(n_d^2 - n_s^2)^2(n_d^4 - n_0^2 n_s^2)}\right)}\right] \quad (2)$$

$$\tan(\Phi_d) \approx \frac{(n_d^2 - n_s^2)^2}{n_s\, k_s \sqrt{(n_0^2 - n_d^2)\,(n_d^2 - n_s^2)}},$$

As with *s*-polarization, the presence of $k_s$ makes only minor corrections to the angle and thickness results. In the limit $k_s \to 0$ we find $\theta_0 \to \tan^{-1}\sqrt{(n_0^2 n_s^2 - n_d^4)/(n_0^2 - n_d^2)^2}$ and $\tan(\Phi_d) \to \infty$. The latter implies that $\Phi_d$ in the lossless case must be equal to $(2m + 1)\pi/2$ for m = 0, 1, 2,.. Now $d = 0$ is no longer a valid solution, so one needs a dielectric layer to get zero reflection even when $k_s = 0$. When $k_s > 0$ the value of $\Phi_d$ is shifted slightly away from these odd multiples of π/2, and hence $\tan(\Phi_d) < \infty$.

**Experimental verification using methyl methacrylate coated silicon substrate**

To experimentally show that a transparent dielectric film on an absorbing substrate exhibits GBA effect in the visible, we spin coated methyl methacrylate (MMA) layer with thickness $t = 500\,nm$ on a silicon (Si) substrate (Supporting Information Figure S1). We measured the reflectance spectra as a function of wavelength (350 nm to 800 nm) and angle of incidence (40° to 85°) using a spectroscopic ellipsometer, see Supporting Information. The false color 2D plot of measured reflectance spectra of *p*- and *s*-polarization is shown in Figure 1a and Figure 1b, respectively. In particular, *p*-polarization exhibits low reflection above incident angle 60° and all angle minimum reflection is obtained for *s*-polarization. However, zero reflection is only possible for a single wavelength and at the Brewster angle for both polarizations. Accordingly, the thin film absorber supports GBA for multiple wavelengths (Supporting Information Figure S2). The rectangulated regions refer to wavelength and angle pairs where the generalized Brewster effect occurs for *p*-polarized light (solid squares) and *s*-polarized light (dashed squares). Clearly, the thin-film absorber supports two modes in the wavelength range of interest for both *p*- and *s*-polarizations.

To further clarify the generalized Brewster effect, we performed reflectance measurements as a function of incident angle by selecting the wavelength in which zero reflection obtained for both polarizations. For *p*-polarization, zero reflection is obtained at 378 nm and 552 nm as shown in Figure 1c and Figure 1d, respectively. The recorded Brewster angle for *p*-polarization at 378 nm and 552 nm is 81° and 76°, respectively. For *s*-polarization, zero reflection is obtained at 450 nm and 752 nm as shown in Figure 1e and Figure 1f, respectively. The obtained Brewster angle for *s*-polarization at 450 nm and 752 nm is 73° and 68°, respectively. The calculated *p*- and *s*-polarization angular reflection was obtained using transfer matrix method (Supporting Information Figure S3). In the model, we solved the Fresnel's equations for a three-layer system (air-MMA-Si) and experimentally obtained refractive indices of MMA were used[26-27]. It is important to note that Brewster angle increases with decreasing the incident wavelength for both polarizations in order to satisfy the amplitude condition for total destructive interference i.e., the amplitude of the out-of-phase partially reflected waves from all interfaces must be equal in magnitude[28]. This is because at lower wavelengths, the reflectance from Si is significantly high. Accordingly, to satisfy the amplitude condition, the reflection from MMA must increase which is only possible at high incidence angles. We have also experimentally investigated the GBA effect in MMA-Ge-glass system and confirmed that this system shows similar Brewster angles, however the incident wavelengths were the effect is observed are slightly red shifted (Supporting Information Figure S4 and Figure S5).

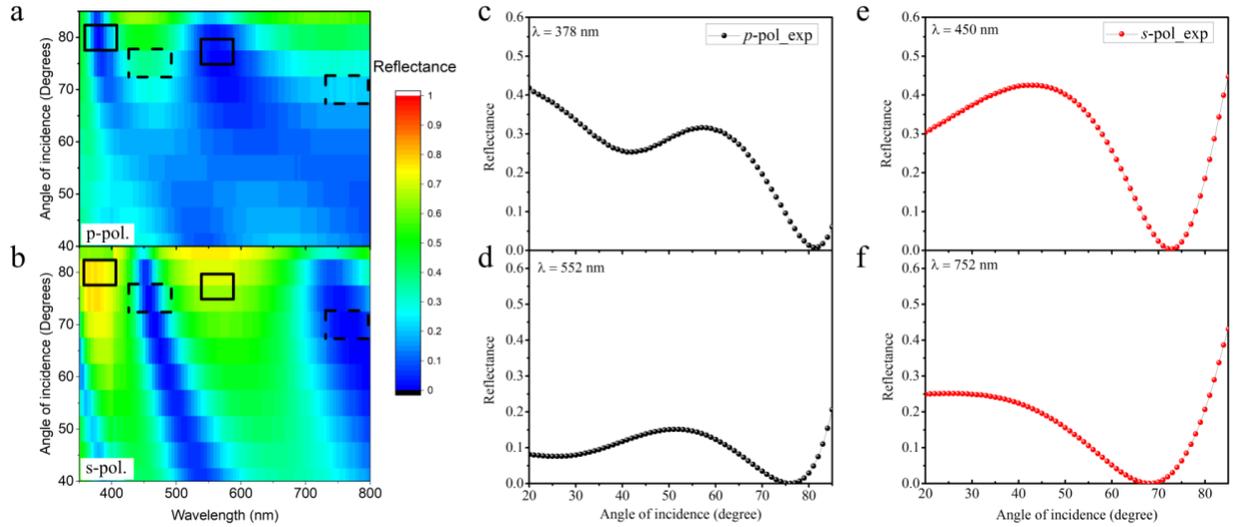

**Figure 1.** Angular reflectivity spectra. (**a**) Measured *p*-polarized and (**b**) *s*-polarized reflectance spectrum. The rectangulated regions refer to wavelength and angle pairs where the generalized Brewster effect occurs for *p*-polarized light (solid squares) and *s*-polarized light (dashed squares). Measured angular reflectance is shown for *p*-polarization at (**c**) 378 nm, and (**d**) 552 nm, and *s*-polarization (c) at 450 nm, and (d) 752 nm.

Note that the system still exhibits the GBA effect even when the lossy substrate is a layer of finite thickness. The mathematical conditions for zero reflectance no longer have tractable analytical forms, but the infinite substrate theory in Eqs. (1-2) remains a reasonable approximation even when the lossy layer thickness is comparable to the wavelength of the incident light. We can see this in Figure 2, which shows incident angle (top row) and dielectric layer thickness d (bottom row) needed to ensure zero reflectance for each polarization. Figure 2 was calculated for an MMA layer of thickness *d* on top of an Si layer of thickness $d_{Si}$ on top of glass and is plotted as a function of $d_{Si}/\lambda$, where $\lambda$ is the incident wavelength. The dashed lines represent the predictions of the theory for infinite Si. One can see that as $d_{Si}$ gets larger the results converge to the theory predictions. The deviations increase as $d_{Si}$ approaches $\lambda$, but within this range are still typically less than 10%.

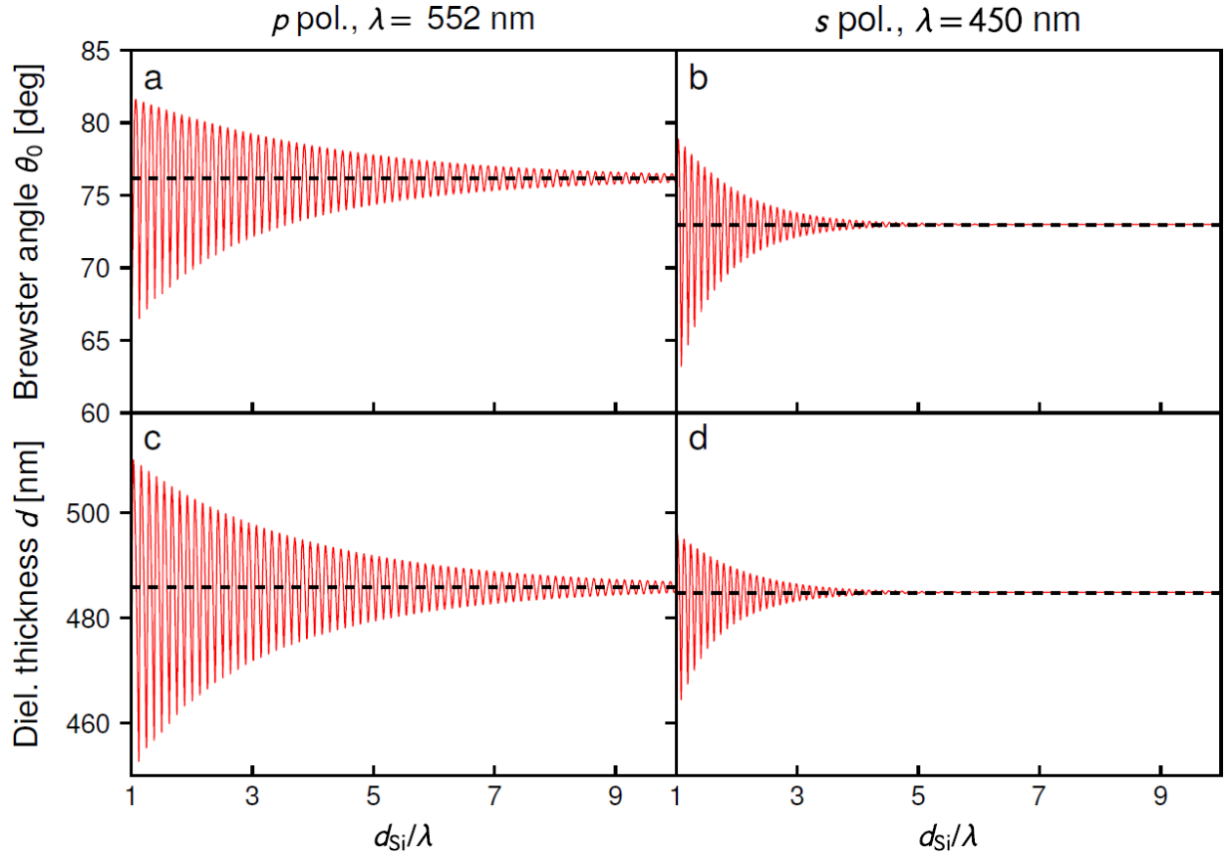

**Figure 2.** Numerical transfer matrix theory results showing the conditions needed to achieve the generalized Brewster angle effect when the lossy substrate layer has a finite thickness. The system consists of: an MMA layer of thickness $d$, an Si layer of thickness $d_{Si}$ on glass. The top row shows the incident angle needed to realize the Brewster effect as a function of $d_{Si}/\lambda$ for **(a)** *p*-polarization, incident wavelength $\lambda = 552$ nm; **(b)** *s*-polarization, incident wavelength $\lambda = 450$ nm. The bottom row **(c-d)** shows the corresponding MMA layer thickness $d$ that is required as an additional condition. The dashed lines correspond to the theoretical predictions for an infinite Si layer, described in Eqs. (1-2) of the main text.

**Experimental verification using light absorbing thin-film metasurface**

Perfect light absorption can occur in ultrathin dielectrics with $t \ll \lambda$, i.e., the dielectric coating does not need to satisfy the anti-reflection coating condition of $t = \lambda/4\,n$. This takes place in a two-layer system with a dielectric coating and a substrate, when the destructive interference phase condition is satisfied due to the existence of an abrupt phase change, i.e., the phase is either 0 or π, at the air/dielectric or the dielectric/substrate interfaces[29]. Because the perfect light absorption in ultrathin films rely on acquiring a phase and not propagating in a refractive medium, these thin-film light absorbers are considered metasurfaces[30].

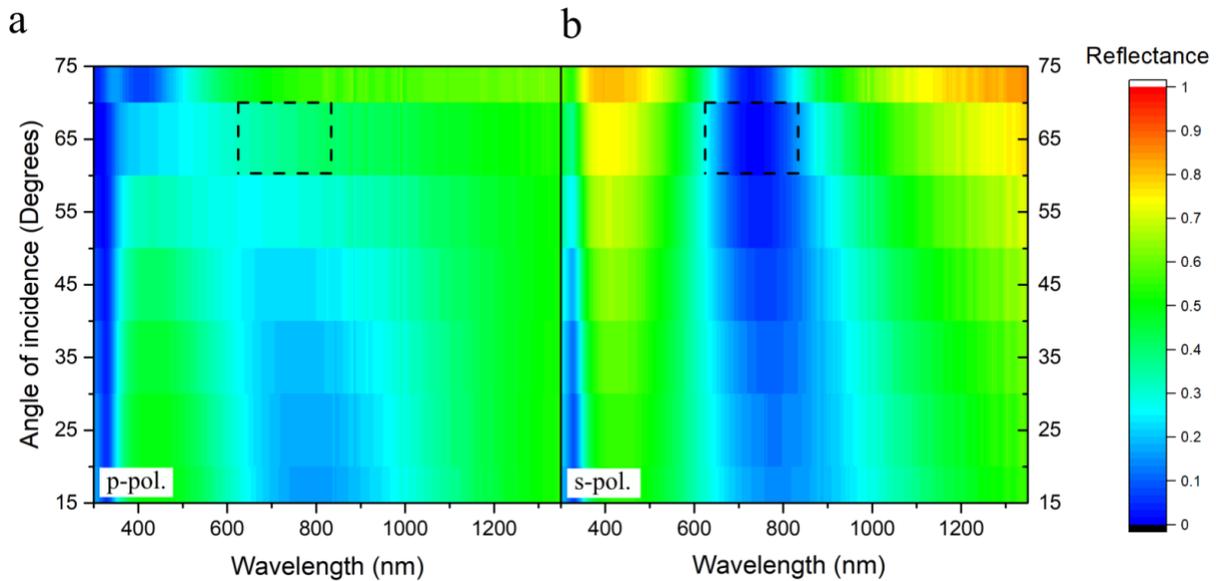

**Figure 3.** Generalized Brewster angle effect (*s*-polarized) using light absorbing thin-film metasurface. Angular reflectivity spectra. (**a**) Measured *p*-polarized and (**b**) *s*-polarized reflectance spectrum.

Figure 3 shows the angular reflectance spectrum for an ultrathin perfect light absorber consisting of a 60 nm TiO$_2$ film on a 100 nm Ni substrate. For *p*-polarized light, perfect light absorption does not occur at any angle or wavelength (Figure 3a). On the other hand, the Brewster angle occur at 750 nm and 68° for *s*-polarized light (Figure 3b). The refractive index of TiO$_2$ at 750 nm is ~ 2.5, i.e., $t = \lambda/5\, n$.

**Experimental verification of singular phase at the GBAs**

Ellipsometry measures the complex reflectance of a system, $\rho$ which is parametrized by the amplitude component $\Psi$ and the phase difference $\Delta$, such that $\tan \Psi = \frac{|r_p|}{|r_s|}$ and $\Delta = \delta_{outgoing} - \delta_{incoming}$, where $\delta$ is the phase difference between the *p*-polarized and *s*-polarized light, such that $\rho = \frac{r_p}{r_s} = \tan \Psi \, e^{i\Delta}$. Accordingly, ellipsometry parameters ($\Psi, \Delta$) have unique characteristics at the Brewster angle. In particular, $\Psi$ reaches a minimum (maximum) at the Brewster angle for *p*-polarized (*s*-polarized) light. Furthermore, beyond the Brewster angle, the reflection phase undergoes ~ $\pi$ phase shift. Accordingly, we can obtain a singular phase (phase difference between *p*- and *s*-polarization) of the reflected light at the zero-reflection wavelength and angle. We show that the lossless dielectric-absorbing substrate system provides singular phase at the Brewster angles of both *p*- and *s*-polarizations.

We experimentally measure the ellipsometry parameters $\Psi$ and $\Delta$ using a variable angle high-resolution spectroscopic ellipsometer. The experimentally obtained $\Psi$ and $\Delta$ spectra of transparent film-absorbing substrate system for wavelengths 378 nm, 552 nm, 450 nm and 752 nm are shown in Figure 4a, 4b, 4c and 4d, respectively. Note that $\psi$ can vary from 0 to 90° and $\Delta$ ranges from 0 to 360° (or -180° to +180°). It is clear that minimum/maximum $\Psi$ value and singular $\Delta$ phase are obtained at the GBAs. For 378 nm and 552 nm wavelengths (Figure 4a and 4b), minimum $\Psi$ is corresponding to Brewster angle for *p*-polarized light. On the other hand, maximum $\Psi$ is obtained at 450 nm and 752 nm (Figure 4c and 4d) corresponding to Brewster angle for *s*-polarized light.

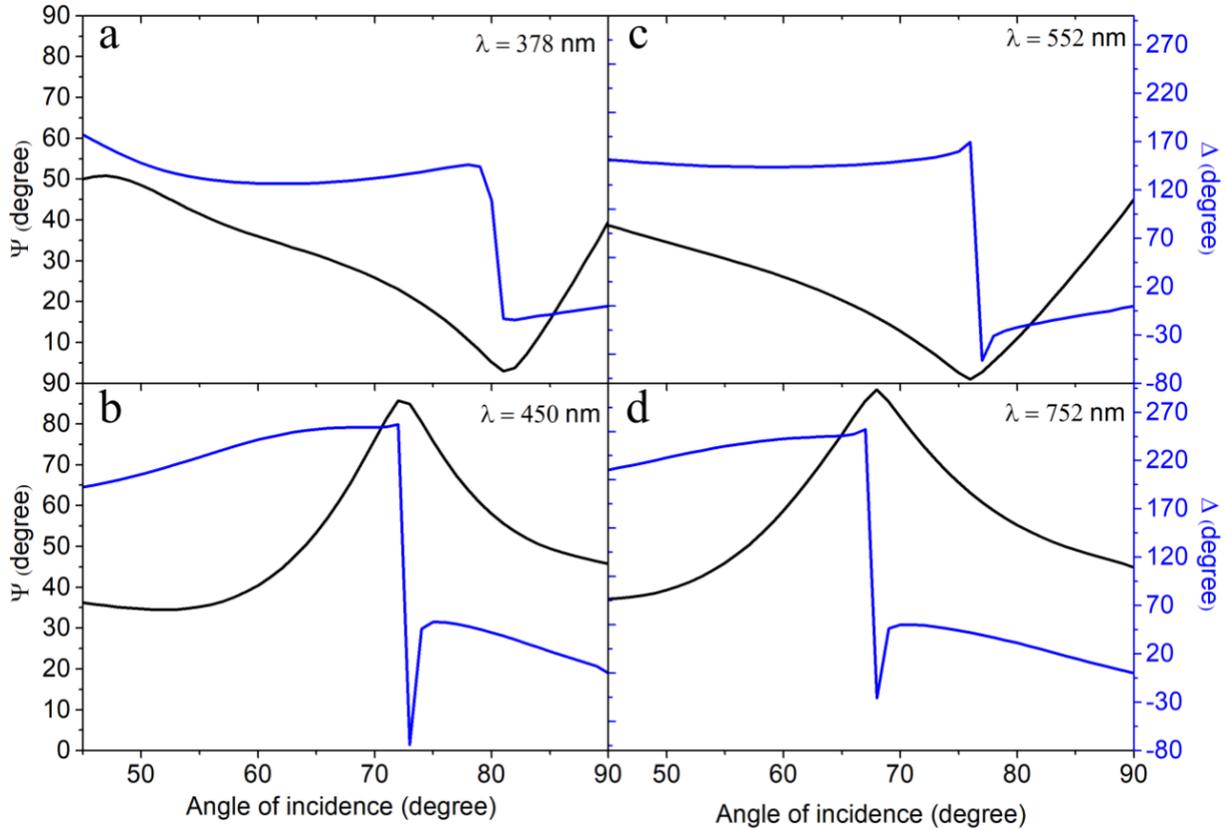

**Figure 4.** Measured pair of ellipsometry parameters ($\psi$ and $\Delta$). (a) at 378 nm, and (b) at 552 nm, (**c**) at 450 nm, and (**d**) at 752 nm. Singular phase is obtained at GBAs in which $\psi$ is minimum/maximum values.

**Hydrogen sensing with single layer graphene transferred on thin-film optical absorbers**

At ambient pressure and temperature, hydrogen is colorless, tasteless and highly flammable[31]. Hydrogen is flammable at concentrations ranging from 4%-75% with low ignition energy. Accordingly, hydrogen sensing is largely used in industries where it is a necessary component or a byproduct to monitor and control the hydrogen partial pressure for safety purposes. Hydrogen is also produced by certain bacteria and hydrogen sensors are used in food industry and have possible medical applications[32-35]. Furthermore, hydrogen sensing is important for fuel cell applications to investigate the loading or unloading kinetics of hydrogen in nanostructured materials.

Various electrical hydrogen sensors based on semiconductors, protonic conductors, and platinum wires have been proposed [31]. However, these systems show enhanced sensitivity only at higher temperatures, which is a major safety issue. In contrast to electrical detection of hydrogen, optical detection techniques offer higher sensitivity in ambient environments, fast response times, and low power consumption. Furthermore, elimination of electric currents and possible sparks in hydrogen rich environments minimizes the risk of explosion. Many approaches for optical hydrogen sensing have been demonstrated using palladium-based optical systems[36-42]. However, most of the palladium-based optical hydrogen sensors reported to date rely on intense lithography techniques.

Here, we exploit the singular phase behavior[4, 26] of our devices at the GBA to detect changes in the optical properties of the thin-film device for hydrogen sensing. Accordingly, we use a transferred graphene layer on the lossless dielectric-lossy substrate light absorber to detect low hydrogen concentrations. The device functions as a lithography-free, large area and inexpensive hydrogen sensor. Graphene is particularly attractive as it can reversibly react with atomic hydrogen[43-44]. Upon hydrogenation, graphene changes its optical properties as it transitions from a semi-metal to an insulator[43]. Furthermore, it was shown that the light-matter interaction of ultra-thin films can enhance drastically based on a strong interference effect in thin-film light absorbers which overcome the limitation between the optical absorption and film thickness[45]. The strong field confinement inside the graphene layer results in ultrahigh sensitivity to the graphene optical properties which we exploit for high sensitivity hydrogen sensing.

Figure 5a is a schematic of the device showing the incoming and outgoing beams undergoing polarization dependent change in amplitude and phase. A CVD grown single layer graphene was transferred on an MMA-Si system using the conventional graphene transfer process. The red curve in Figure 5b shows the Raman spectrum of graphene measured on the fabricated sample. The relative intensity of G and 2D peaks confirm that the transferred graphene is a single layer. Experimentally obtained *p*-polarized reflectance spectrum at 73° incident angle is shown in Figure 5c. Adding a single layer graphene red-shifts the absorption modes, however, perfect light absorption is exclusively realized for *p*-polarized light at 625 nm wavelength and 73° angle of incidence. The red shift in the absorption modes is due to the high complex refractive index of graphene in the visible spectrum[46] (Supporting Information Figure S7 and Figure S8). The addition of graphene further modifies the reflection phase, and ellipsometry parameters of the entire system (Supporting Information Figure S9 and Figure S10). The sensitivity of the mode location on the graphene layer implies that graphene surface chemistry can be effectively studied using the Brewster angle concept. We consider this mode to demonstrate phase-sensitive hydrogen sensing as described below.

The measured ellipsometry parameters $\Psi$ and $\Delta$ of the graphene-MMA-Si system at 625 nm are shown as black curve in Figure 5d and Figure 5e, respectively. One can see that singular phase is obtained at the Brewster angle (73°), where the $\Psi$ value is a minimum. To demonstrate ultra-high sensitivity of the thin-film optical absorber for hydrogen sensing, we used a plasma hydrogenation procedure. In particular, graphene-MMA-Si sample was exposed to different concentrations of atomic hydrogen by controlling the hydrogenation time. As a first step, we studied the Raman spectral features using hydrogenated samples. The hydrogenated graphene shows an additional sharp Raman D peak at about 1340 cm$^{-1}$, which is activated by defects[43-44]. In Figure 5b, we show the emergence of Raman D peak around 1340 cm$^{-1}$ after hydrogenation of the sample (blue curve), which shows the chemical reaction of atomic hydrogen with graphene (Supporting Information Figure S11). On the other hand, the D peak does not exist for graphene with no hydrogenation.

The measured $\psi$ and $\Delta$ spectrum of the sample with different hydrogenation times (1-5 min) are shown in Figure 5d and Figure 5e. The marginal $\psi$ and $\Delta$ shifts at the Brewster angle (73°) with respect to unhydrogenated graphene-MMA-Si sample are shown in Figure 5f. The variation in $\psi$ and $\Delta$ increased with increasing hydrogenation time. However, a drastic change in $\Delta$ is obtained at the Brewster angle compared to $\psi$ change. By considering the phase change obtained for 1 min hydrogenation time (34°) and phase resolution of the instrument (<1°), an areal mass sensitivity of the order of 1 fg/mm$^2$ can be achieved using the proposed platform. Furthermore, we calculated the power dissipation density (W/m$^2$) in the graphene-MMA-Si structure using finite difference time domain (FDTD) method. Figure 5g shows the calculated power dissipation density (W/m$^2$) in the graphene-MMA-Si structure as a function of wavelength. Surprisingly, the power dissipation density is an order of magnitude higher inside the graphene layer compared to the Si substrate. Accordingly, the obtained ultrahigh sensitivity is due to the strong phase sensitivity at the Brewster angle and the strong light-matter interaction at the graphene film. We note here that upon extended hydrogenation, graphene absorption is quenched in the UV, Visible, and IR frequencies[47].

Since the sample required to anneal above 200°C for reversible hydrogenation[43-44], the demonstration of reversible hydrogenation is not possible using the MMA-based thin-film optical absorbers. Nevertheless, the proposed platform can be used to demonstrate reversible hydrogenation by replacing MMA with dielectrics such as SiO$_2$. We show the reversible hydrogenation, however, by measuring the resistance of graphene which indicates its transition from semi-metal to insulator upon hydrogenation, and back to a semi-metal upon annealing (Supporting Information Figure S12).

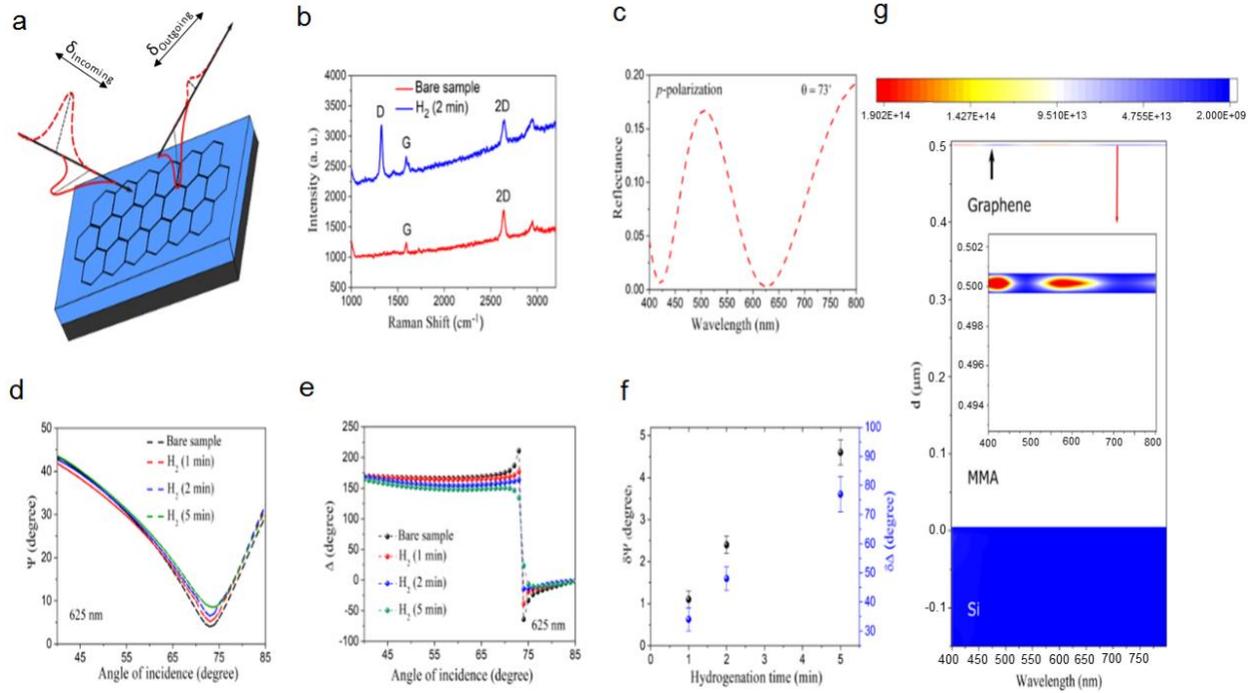

**Figure 5.** Experimental demonstration of hydrogen sensing using graphene-MMA-Si system. (**a**) Schematic of fabricated graphene-MMA-Si system. (**b**) Raman spectrum of bare graphene and hydrogenated graphene acquired from the fabricated structure. The excitation wavelength was 633 nm. (**c**) Measured *p*-polarized reflectance spectrum of graphene-MMA-Si system at 73°. The measured (**d**) $\psi$ and (**e**) $\Delta$ spectrum of graphene-MMA-Si and hydrogenated graphene-MMA-Si for different times at 625 nm. The maximum $\psi$ and $\Delta$ change is obtained at the Brewster angle. (**f**) The marginal $\psi$ and $\Delta$ shifts with different hydrogenation times. (**g**) FDTD calculation of the power dissipation density inside the graphene-MMA-Si structure showing an order of magnitude higher power dissipation in the graphene layer.

In summary, we developed a formalism for generalized Brewster effect in the context of thin-film light absorbers. The experimental demonstrations confirm the existence of a Brewster angle for both *s*- and *p*-polarized light where the light is polarized by reflection due to polarization dependent perfect light absorption. Furthermore, by using an ultrathin lossless dielectric film on a lossy substrate, we realized the Brewster effect using thin-film based metasurface which can act as an effective medium with a tunable Brewster angle. The high phase sensitivity near the GBA, it is possible to monitor slight changes in the electromagnetic environment of the thin-film device which enables a lithography-free ultrasensitive platform. In particular, we showed a hydrogen sensor with readily available and cheap materials, namely, Si, MMA, and graphene without any nanofabrication. The device can also function as a platform for other graphene-based sensors, in particular, for the development of cost-effective apta-biosensor platforms[48]. The effect can be used to realize polarizers for both *s*- and *p*- polarizations, as well as for ultrafast polarization switches[49]. Furthermore, it can be used for the Brewster window in gas lasers, and the optical broadband angular selectivity [14] as well as for Brewster angle microscopy[3].

## ASSOCIATED CONTENT

**Supporting Information**

Sample fabrication and characterizations; angular reflection measurements; spectroscopic ellipsometry characterizations; hydrogenation of graphene-based thin-film absorbers, and additional experimental and simulation results.

## AUTHOR INFORMATION

**Corresponding Authors**

* E-mail: ranjans@ntu.edu.sg (Ranjan Singh)

* E-mail: guo@optics.rochester.edu (Chunlei Guo)

$^$These authors (K. V. Sreekanth and M. ElKabbash) contributed equally to this work.

**Notes**

The authors declare no competing financial interests.

**ACKNOWLEDGMENTS**

The authors (K.V.S. and R.S.) acknowledge Singapore Ministry of Education (MOE) Grant No. MOE2015-T2-2-103 for funding of this research.

# Supporting Information: Generalized Brewster-angle effect in thin-film optical absorbers and its application for graphene hydrogen sensing

**Methods and Materials**

**Sample fabrication:**

Thin-film absorbers: To fabricate the thin film absorber, 500 nm thick MMA copolymer from MICROCHEM (8.5MMAEL 11) was spin coated at 3000 rpm on 700 µm thick Si substrates and 30 nm thick Ge layer deposited glass substrates. The Ge layer was deposited by thermal evaporation (Oerlikon Leopold vacuum system) of Ge pellets at a deposition rate of 0.2 Å/s and a base pressure of <5 x $10^{-6}$ mbar.

Thin-film meta-surface light absorber: Films were deposited on a glass substrate (Micro slides, Corning) using electron-beam (e-beam) evaporation of Ni and $TiO_2$ pellets (Kurt J. Lesker). The deposition rate of Ni and TiO2 were 0.5 and 1 10 A. $s^{-1}$.

Graphene-based thin-film absorbers: To fabricate the graphene based thin film absorber, the graphene was grown on 25 µm thick copper foil using the conventional chemical vapor deposition method. Prior to deposition of graphene, copper foil was cleaned thermally at 1000 °C in the presence of 2 SCCM hydrogen. Thereafter, 10 SCCM $CH_4$ injected into the chamber to deposit the graphene at 1000 °C for one hour. After deposition of graphene, the copper foil was rapidly cool down to room temperature. 500 nm thick MMA was spin coated on the graphene/Cu at 3000 rpm. MMA/graphene deposited on copper foil was transfer to the Si substrates using the wet chemical etching technique. The MMA/graphene/Cu was immersed in the $FeCl_3$ solution with its MMA/graphene facing upward for 2 hours. After copper etching, remaining MMA/graphene, was transferred to the deionized water to remove the unwanted impurities from the $FeCl_3$ solution. The cleaning process was repeated for 3 times with 1-liter fresh deionized water. The cleaned MMA/graphene was transfer to the Si to design the graphene/MMA/Si hetero-structure. Thereafter, the designed graphene/MMA/Si was kept in the 60 °C preheated oven for 6 hours to evaporate the deionized water. The designed structure was used for the hydrogenation using plasma.

**Angular reflection measurements:** Angular reflection was measured using Variable-angle high-resolution spectroscopic ellipsometer (J. A. Woollam Co., Inc, V-VASE). The transmittance is zero for all wavelengths and angles. Since we are dealing with thin films, perfect light absorption corresponds to near zero reflectance. We consider $R \leq 5 \times 10^{-4}$ to be zero reflectance since this is the noise limit of the detector which is determined by considering an incident $p$-polarized light on a non-chiral film and measuring the $s$-to-$p$ reflectance which gives us the detector noise level for the parameters adopted in our measurements.

**Hydrogenation of graphene-based thin-film absorbers:** Hydrogenation of graphene transferred on the graphene/MMA/Si were performed low power capacitively coupled RF plasma of the hydrogen and argon gas. The graphene/MMA/Si hetero-structure was place in the quartz tube and evacuated to $10^{-3}$ mbar prior to the hydrogenation. To perform hydrogenation, RF plasma was generated between the two copper electrodes on the quartz tube. To this end, the 30-Watt RF plasma was generated by flowing 2 SCCM hydrogen and 15 SCCM argon gas while maintaining the chamber pressure of 0.13 mbar. The sample was placed in between the electrodes separated by a safe distance of 15 cm from the discharge zone to avoid the direct exposure. The graphene/MMA/silicon was exposed to three different time duration of RF plasma i.e. 1, 2 and 5 minutes. The plasma exposure will attach the hydrogen to the surface of the graphene. The hydrogenation of graphene in the graphene/MMA/Si structure was investigated by measuring the intensities of D and G peak of the hydrogenated graphene as a function of the plasma exposure using the Raman spectra (Renishaw Invia spectrometer, 632.8 nm excitation wavelength). The estimated distance between the defects after first hydrogenation (for 1 min) is, $L_D \approx 20 nm$ which is far below the low-defect density regime[1].

**Spectroscopic characterizations:** Variable-angle high-resolution spectroscopic ellipsometer (J. A. Woollam Co., Inc, V-VASE) was used to determine the thicknesses and optical constants of MMA and Ge layers. The polarized reflectance spectra as a function of wavelength and incident angle were acquired using the same instrument with a wavelength spectroscopic resolution of 2 nm and an angular resolution of 1°. The ellipsometry parameters ($\psi$ and $\Delta$) as a function of incident angle were acquired using the VASE ellipsometer with an angular resolution of 1°. All the $\psi$ and $\Delta$ spectra were acquired by selecting high accuracy mode in the ellipsometer. Furthermore, the complex refractive indices of MMA, Ni, and TiO$_2$ were obtained by fitting the ellipsometry parameters.

**Numerical simulations:** Numerical reflection spectra and ellipsometry parameters spectra were generated using a transfer matrix method-based simulation model written in Matlab. The calculated reflectance spectra of Graphene-MMA-Si device and the calculated power dissipation distribution in the thin-film stack was performed using the commercially available finite-difference time-domain software from Lumerical®. The simulation was performed using a 2D model with incident plane wave at zero incidence angle. Periodic boundary conditions were used in the x-direction and perfectly matched layers where used in the y-direction (normal to the sample). The mesh was tailored to each layer with a mesh step of 0.0005 µm for the graphene layer and 0.005 µm for the rest of the structure.

**Theoretical framework**

We investigate the proposed design, i.e., a lossless dielectric film on a substrate with optical losses. Our system consists of a superstrate (refractive index $n_0$), a dielectric layer (refractive index $n_d$, thickness $d$), and a lossy substrate (refractive index $n_s + ik_s$). Using transfer matrix theory[2], one can express the complex reflection coefficient $r$ of this system as

$$r = \frac{\gamma_0 M_{11} + \gamma_0 \gamma_s M_{12} - M_{21} - \gamma_s M_{22}}{\gamma_0 M_{11} + \gamma_0 \gamma_s M_{12} + M_{21} + \gamma_s M_{22}} \quad (1)$$

where $M_{ij}$ are the entries of the transfer matrix $M$ for the dielectric layer,

$$M = \begin{pmatrix} \cos(\Phi_d) & -i\gamma_d^{-1}\sin(\Phi_d) \\ -i\gamma_d \sin(\Phi_d) & \cos(\Phi_d) \end{pmatrix} \quad (2)$$

In Eqs. (1) - (2) the factors $\gamma_l$ for each layer $l = 0, d, s$ are given by

$$\gamma_0 = \begin{cases} n_0^{-1}\sqrt{1 - \sin^2(\theta_0)} & p\ pol. \\ n_0 \sqrt{1 - \sin^2(\theta_0)} & s\ pol. \end{cases}, \gamma_d = \begin{cases} n_d^{-1}\sqrt{1 - \dfrac{n_0^2 \sin^2(\theta_0)}{n_d^2}} & p\ pol. \\ n_d \sqrt{1 - \dfrac{n_0^2 \sin^2(\theta_0)}{n_d^2}} & s\ pol. \end{cases}, \quad (3)$$

$$\gamma_s = \begin{cases} (n_s + ik_s)^{-1}\sqrt{1 - \dfrac{n_0^2 \sin^2(\theta_0)}{(n_s + ik_s)^2}} & p\ pol. \\ (n_s + ik_s)\sqrt{1 - \dfrac{n_0^2 \sin^2(\theta_0)}{(n_s + ik_s)^2}} & s\ pol. \end{cases}$$

where $\theta_0$ is the incident angle and the two results in each case correspond to $p$ and $s$ polarized incident light. The angle $\Phi_d \equiv 2\pi d \lambda^{-1}\sqrt{n_d^2 - n_0^2 \sin^2(\theta_0)}$ in Eq. (2) is the phase thickness of the dielectric layer, where $\lambda$ is the wavelength of the incident light. Note that $\gamma_0$ and $\gamma_d$ are both real, while $\gamma_s = \gamma_s^R + i\gamma_s^I$ is complex when $k_s > 0$, with real and imaginary parts $\gamma_s^R$ and $\gamma_s^I$, respectively.

Zero reflectance occurs when the numerator of Eq. (1) is equal to zero, which is equivalent to requiring that the real and imaginary parts of the numerator separately equal zero. This gives two equations defining the conditions for zero reflectance,

$$\gamma_d(\gamma_0 - \gamma_s^R) + \gamma_0 \gamma_s^I \tan(\Phi_d) = 0, \qquad (4)$$

$$\gamma_d(\gamma_s^I - \gamma_d \tan(\Phi_d)) + \gamma_0 \gamma_s^R \tan(\Phi_d) = 0.$$

Exact analytical solutions for the incident angle $\theta_0$ and the thickness $d$ that satisfy Eq. (4) are not possible to find when $k_s > 0$. However, by Taylor expanding the expressions in Eqs. (3-4) for small $k_s$, one can find the forms of $\theta_0$ and $\tan(\Phi_d)$ that satisfy Eq. (4) to leading order in $k_s$. These are shown for *p*- polarization as Eq. (1) in the main text, and for *s*- polarization as Eq. (2) in the main text. Numerically, the approximate form for the zero-reflectance condition turns out to be a good approximation for both the Si and Ge substrates considered in the experiments, without the necessity of including correction terms at higher order in $k_s$.

The ellipsometry parameters of air-MMA-Si system were calculated by using TMM simulation model. The $\psi$ and $\Delta$ were obtained from, $\tan\Psi = \left|\dfrac{r_p}{r_s}\right|$ and $\tan\Psi \exp(i\Delta) = \dfrac{r_p}{r_s}$. In order to obtain good match with experimental results, the thickness of MMA layer was slightly varied from the measured value. The MMA layer thicknesses used in the simulation were: 496 nm at 378 nm, 484 nm for 450 nm, 483 nm for 552 nm and 483 nm for 752 nm.

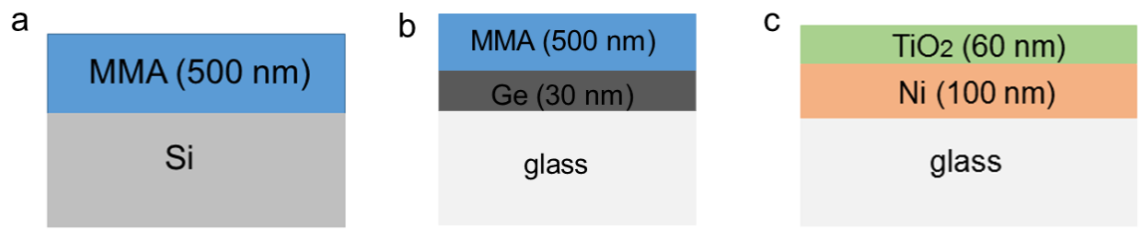

Figure S1. Schematic of fabricated thin-film optical absorber samples. (a) MMA-Si system, (b) MMA-Ge-glass system and (c) TiO$_2$-Ni-glass system.

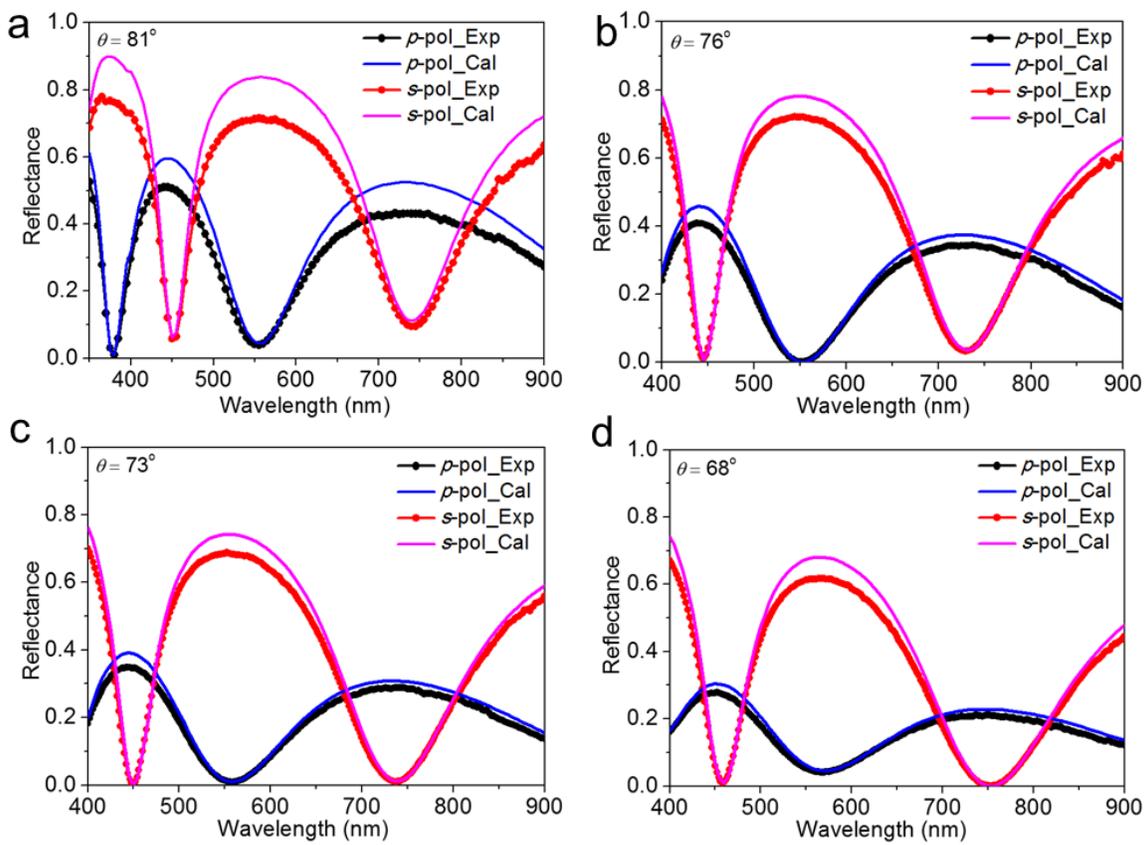

Figure S2. Experimental and calculated reflectance spectrum of MMA-Si system for *p*- and *s*-polarizations. (a) $\theta$=81°, (b) $\theta$=76°, (c) $\theta$=73° and (d) $\theta$=68°.

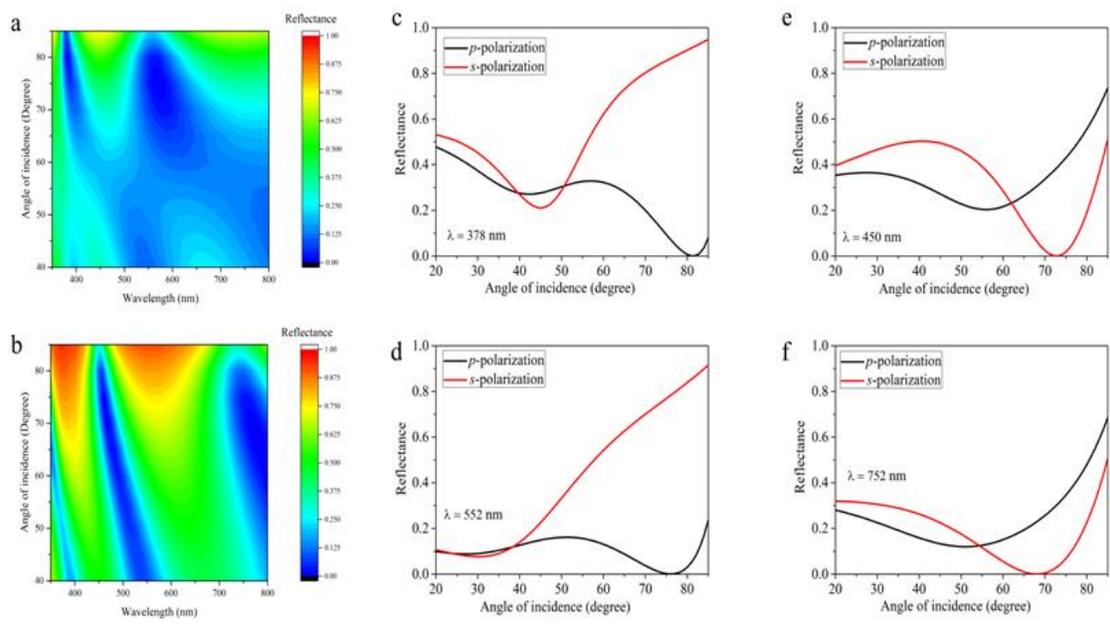

Figure S3. Calculated reflectance spectrum of MMA-Si system for *p*- and *s*-polarizations. 2D map of reflectance spectrum for (a) *p*-polarization and (b) *s*-polarization. Calculated angular reflectance is shown for both *p*- and *s*- polarizations at (c) 378 nm, (d) 552 nm, (e) 450 nm, and (f) 752 nm.

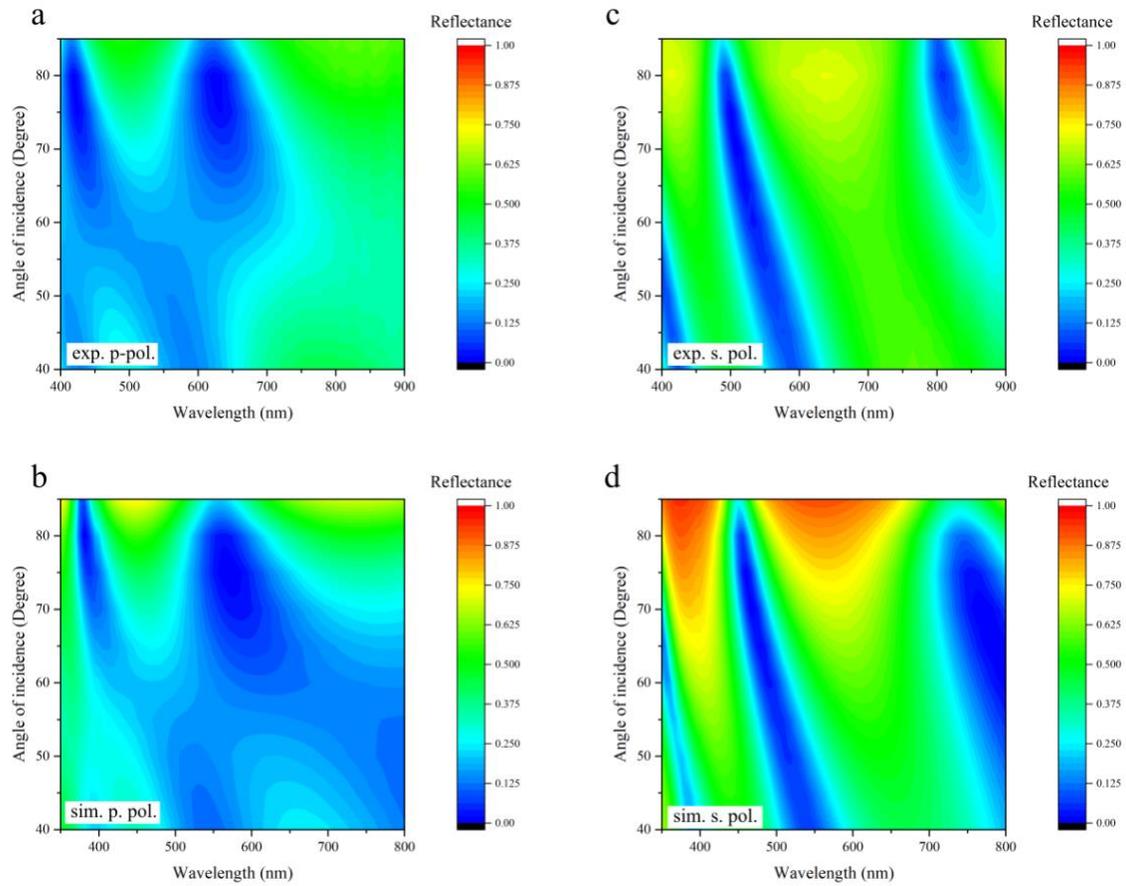

Figure S4. 2D map of reflectance spectra as a function of angle of incidence for MMA-Ge-glass system. For *p*-polarization (a) Experiment and (b) Calculated. For *s*-polarization (c) Experiment and (d) Calculated.

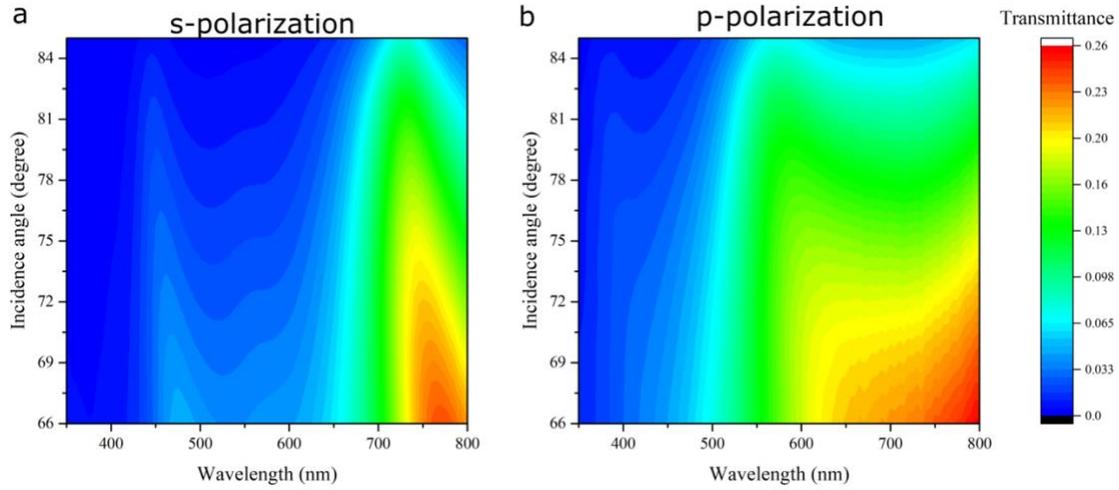

Figure S5. 2D map of experimentally measured transmittance spectra as a function of angle of incidence for MMA-Ge-glass system for (a) *s*-polarized light, and (b) *p*-polarized light.

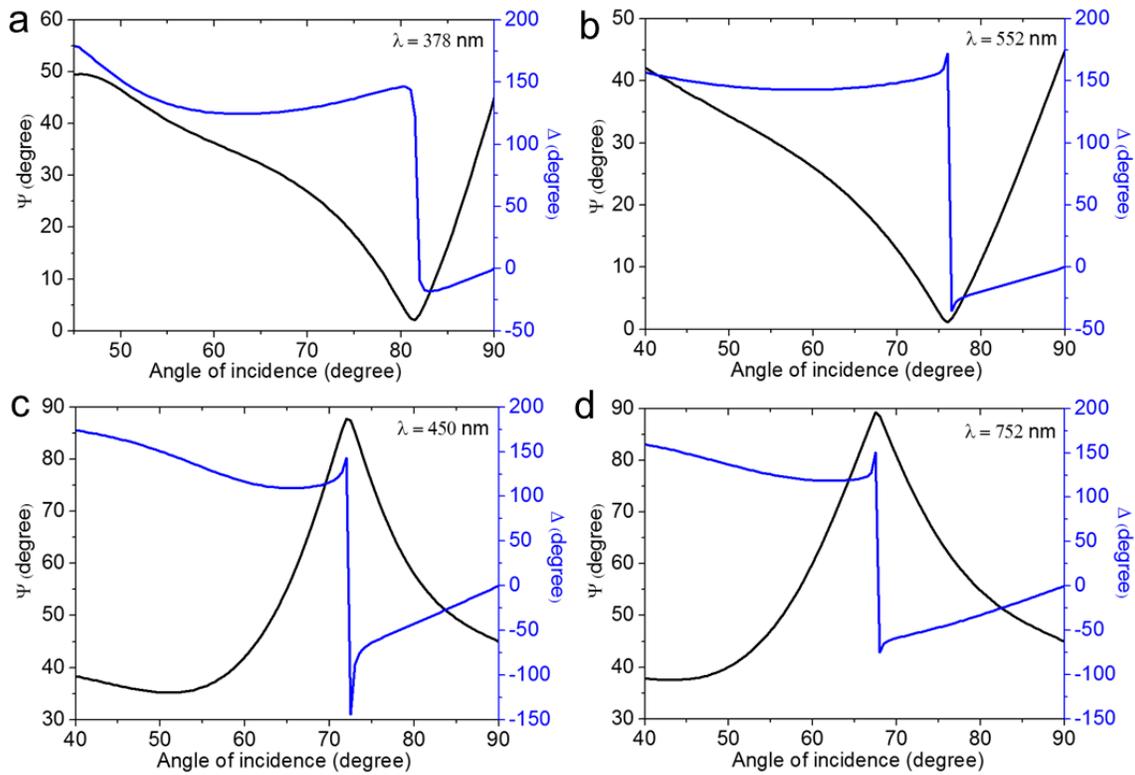

Figure S6. Calculated ellipsometry parameters ($\psi$ and $\Delta$) of MMA-Si system. (a) at $\lambda$=378 nm, (b) at $\lambda$=552 nm, (c) at $\lambda$=450 nm and (d) at $\lambda$=752 nm.

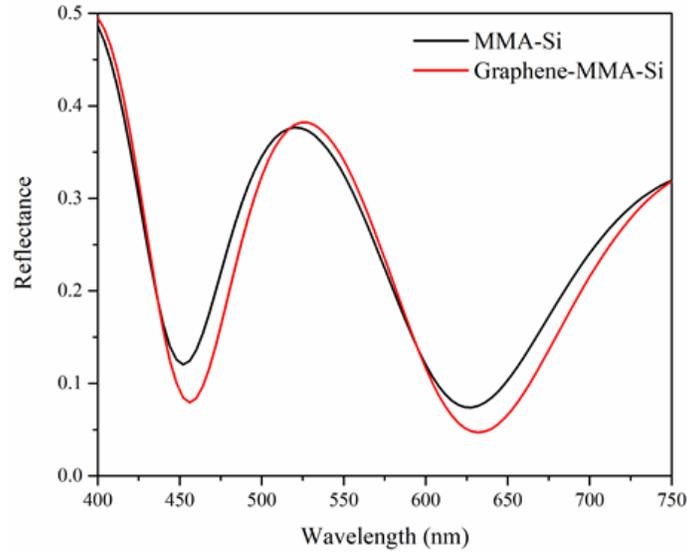

Figure S7. Calculated reflectance spectrum of graphene-MMA-Si and MMA-Si at normal (0º) incidence showing the effect of adding a graphene layer on light absorption.

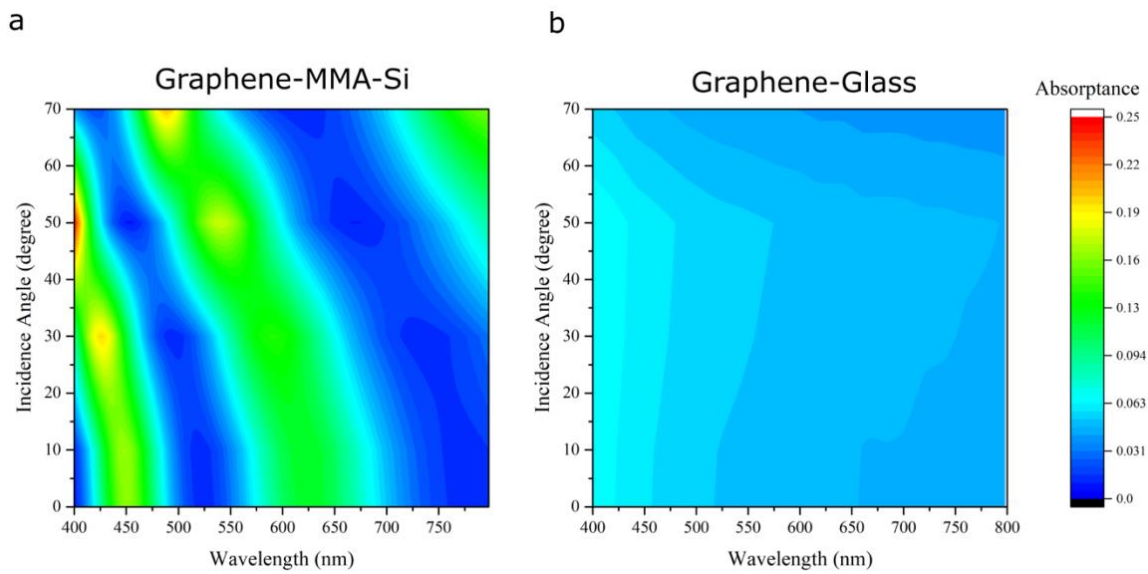

Figure S8. The isolated absorptance of graphene layer calculated on an MMA-Si stack (a) and of graphene on glass (b) which shows the enhanced absorptance of graphene due to destructive interference.

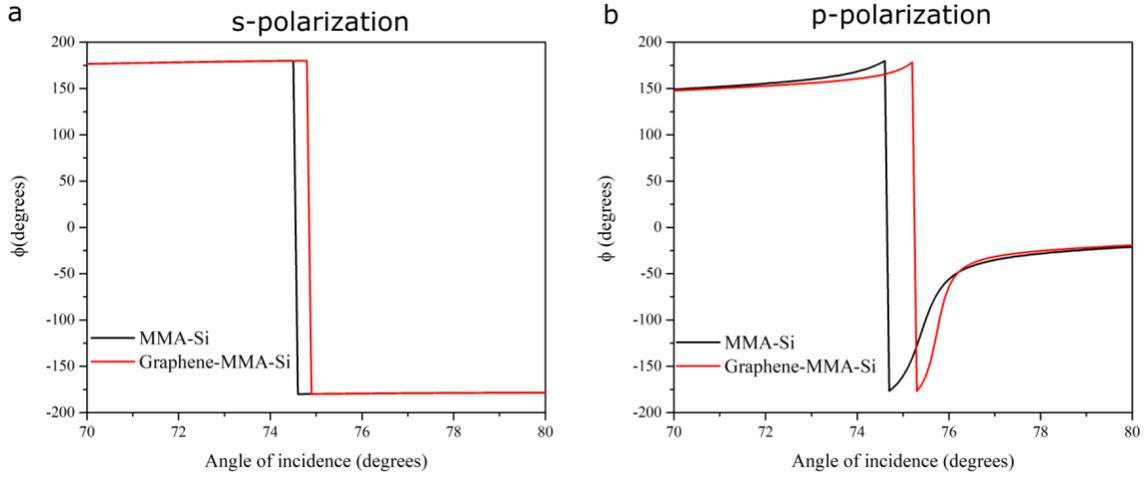

Figure S9. The calculated phase upon reflection from an MMA-Si stack vs. Graphene-MMA-Si stack as a function of angle for (a) *s*-polarized light and (b) *p*-polarized light.

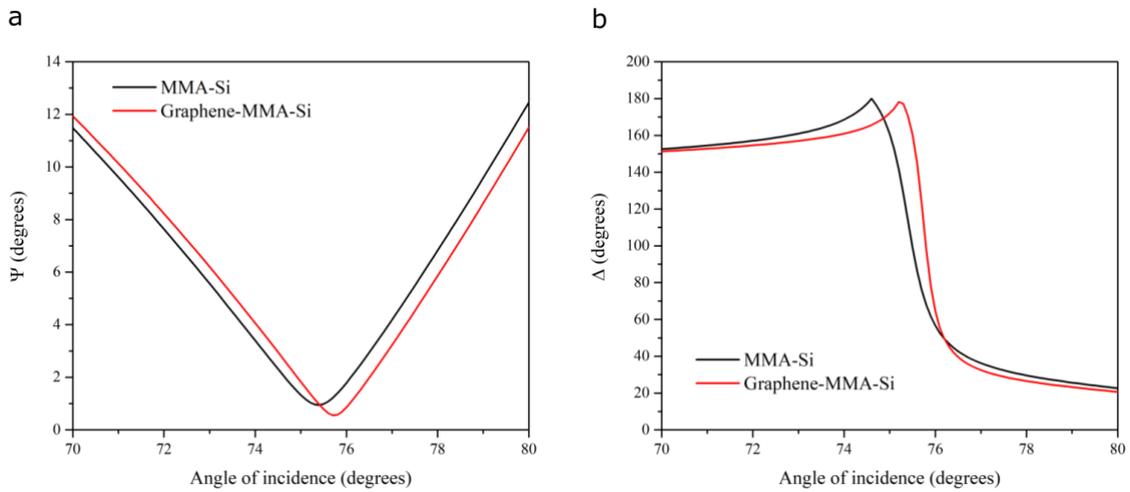

Figure S10. The calculated (a) ψ (b) Δ spectra at the Brewster angle for an MMA-Si stack vs. Graphene-MMA-Si. Both ψ and Δ values, which determine the Brewster angle, change by adding graphene layer.

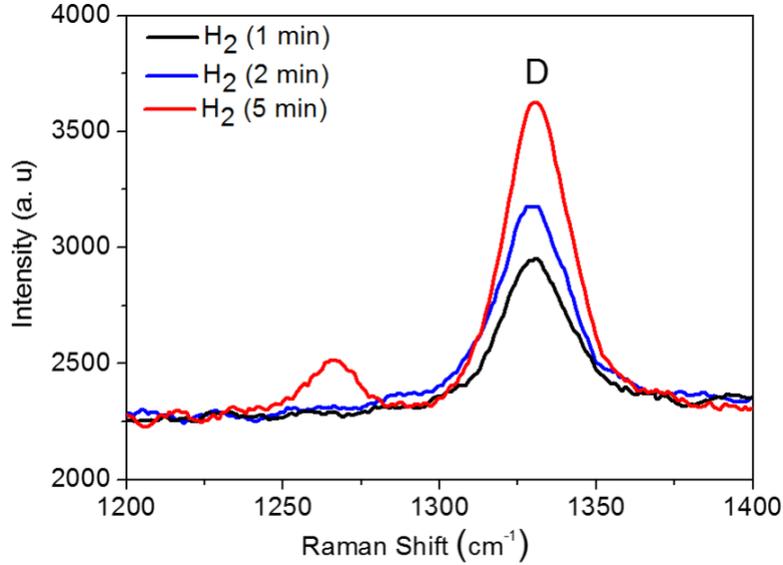

Figure S11. Raman D peak change with hydrogenation time.

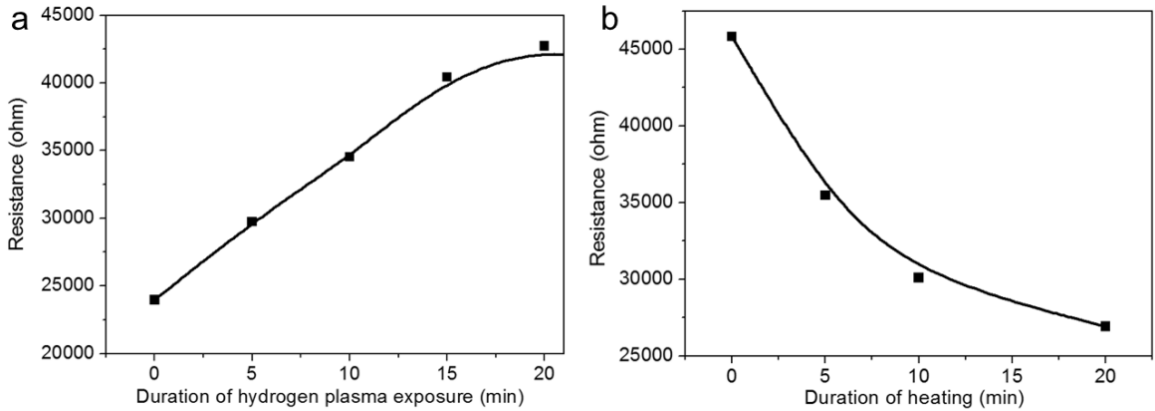

Figure S12. Reversible hydrogenation of Graphene-MMA-Si stack: As graphene is hydrogenated, it transitions from a semi-metal to an insulator. Accordingly, hydrogenation is associated with increase in the measured resistance across the graphene sheet as we see in (a). As we anneal the sample at 150°C, the measured resistance drops as shown in (b) indicating that graphene is becoming a semi-metal. Note that here we used higher plasma exposure time because the sample was placed in between the electrodes separated by a distance of 20 cm.